\documentclass{PoS}
\usepackage{amsmath}
\usepackage{amssymb}
\usepackage{graphicx}

\title{New Mechanism for Neutrino Mass Generation and Triply Charged Higgs Boson at the LHC}

\ShortTitle{Neutrino Mass Generation}

\author{\speaker{S. Nandi}\thanks{OSU-HEP-10-10}\\
        Department of Physics,Oklahoma State University
        and Oklahoma Center for High Energy Physics,
        Stillwater, OK 74078, USA\\
        E-mail: \email{s.nandi@okstate.edu}}

\abstract{ In this talk, I present a new mechanism for the
generation of neutrino masses via dimension 7 operators:
$l~\!l~\!H~\!H(H^{\dagger}H)/M^3$. This leads to new formula for the
light neutrino masses, $m_{\nu} \sim v^4/M^3$. This is distinct from
the usual see-saw formulae :  $m_{\nu} \sim v^2/M$. The scale of new
physics can naturally be at the TeV scale. Microscopic theory that
generated $d=7$ operator has an isospin 3/2 Higgs multiplet $\Phi$
containing a triply charged Higgs boson with mass around
$\sim\textrm{TeV}$ or less.  These particles can be produced at the
LHC (and possibly at the Tevatron) with distinctive multi-$W$ and
multi-lepton final states.  For some choice of the parameter space,
these particles can also be long-lived with the possibility of
displaced vertices, or even escaping the detector. Their leptonic
decay modes carry information about the nature of the neutrino mass
hierarchy.
}

\FullConference{35th International Conference of High Energy Physics - ICHEP2010,\\
        July 22-28, 2010\\
        Paris France}

\begin{document}

\section{Introduction}\vspace{-0.25cm}
The existence of neutrino masses is now firmly established. This is the
first indication for physics beyond the SM; other indications being the existence of dark matter and baryon asymmetry of the universe.  The neutrino mass $m_{\nu}\sim
10^{-2}\textrm{ eV}$ is about a billion times smaller than the quark and
charged lepton masses.  What is the mechanism for such a tiny
neutrino mass generation?  The most popular mechanism for generating
neutrino masses is the see-saw mechanism \cite{seesaw}, $m_{\nu}\sim m^2_D/M$. The
corresponding effective interaction in the SM is the dimension 5
operator: $\mathcal{L}_{\textrm{eff}} = (f/M)~\!l~\!l~\!H~\!H$.
This implies a new symmetry breaking scale $M$.  This scale is too
high and no connection can be made to the physics to be explored at
the LHC or Tevatron.  It is important to explore alternate mechanisms \cite{loop,gn,zurab} which can be more directly tested.

It is possible that the dimension $5$ operator does not contribute to
neutrino masses in a significant way.  The next operator (dimension
7) is $\mathcal{L}_{\textrm{eff}} =(f/M^3)~\!l~\!l~\!H~\!H
(H^{\dag}H)$. This by itself is not enough to make $M
\sim\textrm{TeV}$ because it requires $f \sim 10^{-9}$.  We propose
a model in which $f \sim y_1 y_2 \lambda_4$ with each factor $\sim
10^{-3}$ (domain of natural values).  This gives $M
\sim\textrm{TeV}$ for neutrino masses in the range $10^{-2} -
10^{-1}\textrm{ eV}$.  This will connect the neutrino physics to the
physics being explored at the LHC and Tevatron.

\vspace{-0.25cm}
\section{Model and the formalism}\vspace{-0.25cm}

The gauge symmetry in our model is the $SM = SU(3)_c \times SU(2)_L
\times U(1)_Y $. In addition to the usual SM fields, there is a pair
of vector-like $SU(2)_L$ triplet leptons $\Sigma + \bar{\Sigma}$
transforming as $(1,3,2)$ and $(1,3,-2)$ where $\Sigma =
(\Sigma^{++}, \Sigma^+, \Sigma^0)$, and a new isospin $3/2$ Higgs
$\Phi$ with components $(\Phi^{+++}, \Phi^{++}, \Phi^+, \Phi^0)$.
The $\Phi$ has positive mass squared term, but acquires a tiny VEV
through interactions with $H$.  The $\Sigma$ has interactions with
the SM lepton doublets, $H$, and $\Phi$.

The Higgs potential in our model is given by
\begin{align*}
    V=&-\mu^2_H~ H^{\dag}H+M^2_{\Phi}~\Phi^{\dag}\Phi
      +\lambda(H^{\dag}H)^{2}+\lambda_1(\Phi^{\dag}\Phi)^{2}
      +\lambda_2(H^{\dag}H)(\Phi^{\dag}\Phi)\\
      &+\lambda_3(H^{\dag}\frac{t_a}{2}H)(\Phi^{\dag}\frac{T_A}{2}\Phi)
      +\lambda_4 (HHH\Phi + \Phi^{\dag} H^{\dag}H^{\dag}H^{\dag}).
\end{align*}
Minimization of $V$ gives $\langle\Phi_0\rangle \equiv v_{\Phi}\sim
-\lambda_4 v^3_H/M^2_{\Phi}$.

\textbf{Light neutrino mass generation:}  The light neutrino masses
are generated by combining the following interactions:
 $\mathcal{L}=y_il_i H^*\Sigma +\bar{y}_il_i\Phi\bar{\Sigma
}+M_{\Sigma }\Sigma\bar{\Sigma }$ where $y_i$, $\bar{y}_i$ are
dimensionless Yukawa couplings.  This gives rise to an effective
dimension 7 interaction (see Fig.~1),
\begin{align*}
\mathcal{L}_{eff} &=
\frac{(y_i\bar{y}_j+y_j\bar{y}_i)}{M_\Sigma}l_il_j H^*\Phi +h.c.,&
\textrm{with }  v_{\Phi} &= -\lambda_4 \frac{v^3_H}{M_{\Phi}^2},&
\textrm{ and with } (y_1, y_2,\lambda_4) &\sim 10^{-3}.
\end{align*}
\begin{figure}
\vspace{-0.75cm}
\begin{center}
\begin{tabular*}{\textwidth}{@{}p{0.5\textwidth}@{}@{}p{0.5\textwidth}@{}}
    \includegraphics[width=0.4\textwidth]{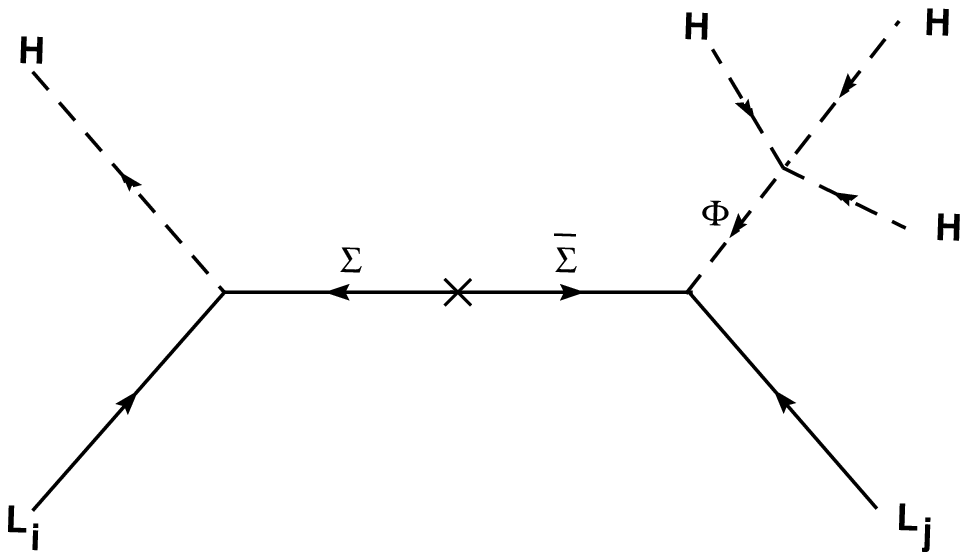}\caption{Effective 7 dimensional interaction.}&
    \includegraphics[width=0.5\textwidth]{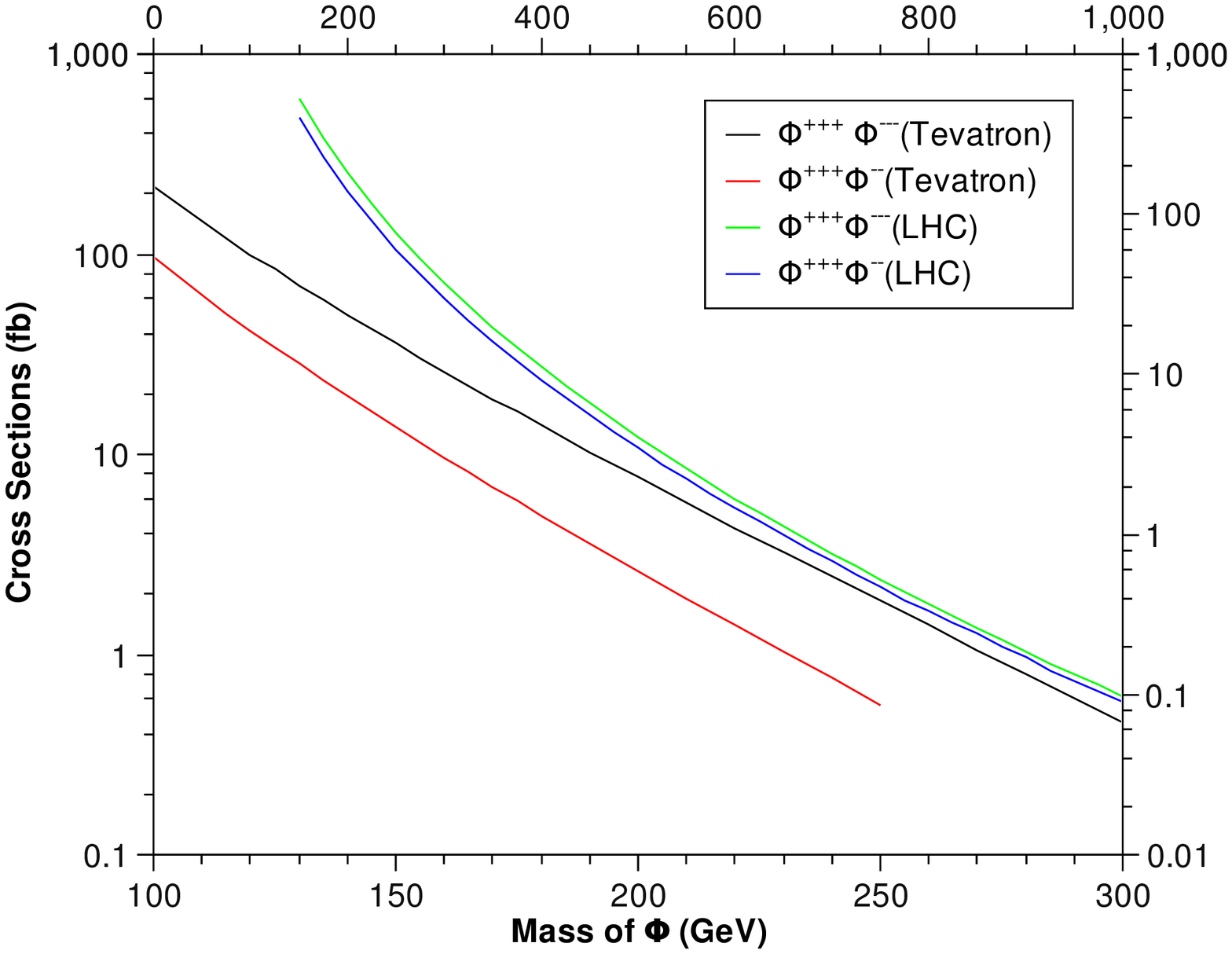}\caption{Production cross sections. Use top and right axes for LHC.  Use bottom and left axes for Tevatron.}
\end{tabular*}
\end{center}
\vspace{-0.75cm}
\end{figure}
\textbf{Mass spectrum of $\Phi$:}  The mass spectrum of $\Phi$ is
given by $M^2_{\Phi_i} = M^2_{\Phi} + \lambda_2 v^2_H - \frac{1}{2}
\lambda_3 I_{3i} v^2$, where $I_{3i} = (3/2, 1/2, -1/2, -3/2)$ for
$(\Phi^{+++},\Phi^{++},\Phi^+,\Phi^0)$ respectively. 
The two possible hierarchies for the spectrum of $\Phi$ are\\
\begin{tabular*}{\textwidth}{lcr}
Positive $\lambda_3: M_{\Phi^{+++}}<M_{\Phi^{++}}<M_{\Phi^{+}}<M_{\Phi^{o}}$&
\hfill{ }&
Negative $\lambda_3: M_{\Phi^{+++}}>M_{\Phi^{++}}>M_{\Phi^{+}}>M_{\Phi^{o}}$.
\end{tabular*}

\textbf{Parameters and existing constraints:} The model parameters
are $v_{\Phi}$, $M_{\Phi}$, $M_{\Sigma}$, and $\Delta M$.

The $\Phi$ has isospin 3/2 and contributes to the $\rho$ parameter
at tree level, $\rho = 1-(6 v^2_{\Phi}/v^2_H)$.  The experimental
value is $\rho = 1.0000^{+0.0011}_{-0.0007}$ \cite{Amsler:2008zzb}.  At the $3\sigma$
level we get $v_{\Phi}<2.5\textrm{ GeV}$. The mass splittings between the
components of $\Phi$ induces an additional positive contribution to
$\rho$ at one loop level, $\Delta \rho \simeq (5 \alpha_2/6 \pi)
(\Delta M/m_W)^2$, thus $\Delta M < 38\textrm{ GeV}$.

%

\textbf{Experimental constraints:}  A charged $\Phi$ has a mass
bound from LEP2 to be $>100\textrm{ GeV}$ \cite{lep}.  The CDF and D0
Collaborations have looked for stable CHAMPS (charged massive
particle).  Using CDF cross sections times branching ratio limits,
we obtain mass $>120\textrm{ GeV}$ for stable, charged $\Phi^{+++}$ \cite{tevatron}.

\textbf{Productions:}  In hadronic collisions, the triply charged
Higgs bosons can be pair produced via the Drell-Yan process.  Their
cross sections at the Tevatron and the LHC (at 14 TeV) are shown in
Fig.~2, where $pp$ or $p \bar{p}\rightarrow \Phi^{+++} \Phi^{---}
\rightarrow 6 W$, $4 W l^+ l^+ $, $4 W l^- l^- $, or $2W l^+ l^+ l^-
l^-$ with or without displaced vertices, depending on $v_\Phi$.

\vspace{-0.25cm}
\section{Phenomenological Implications}\vspace{-0.25cm}

The two possible mass hierarchies of $\Phi$ have $\Phi^{+++}$ as the
lightest or heaviest. We consider the case in which $\Phi^{+++}$ is
the lightest.  The phenomenological implications are most
distinctive with displaced vertices for this case.

\textbf{Decays:}  There are two possible decay modes.  The
$\Phi^{+++} \rightarrow W^{+} W^{+} W^{+}$ mode dominates for
higher values of $v_{\Phi}$.  The $\Phi^{+++}\rightarrow
W^{+}l^{+}l^{+}$ mode dominates for smaller values of $v_{\Phi}$.

%
\begin{figure}
\vspace{-0.75cm}
\begin{center}
\begin{tabular*}{\textwidth}{@{}p{0.5\textwidth}@{}@{}p{0.5\textwidth}@{}}
        \includegraphics[width=0.55\textwidth]{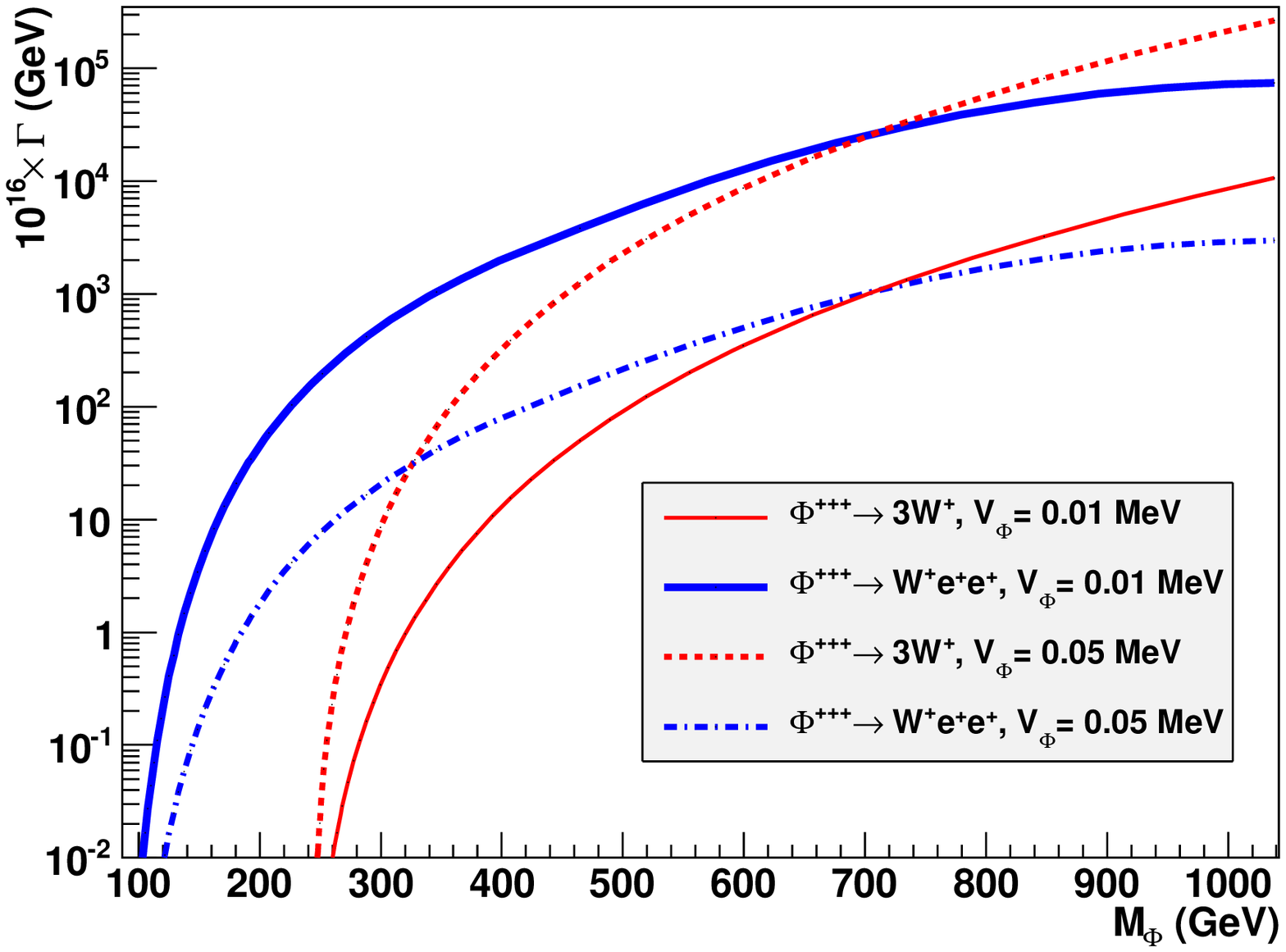}\caption{Decay widths as functions of $M_\Phi$.}&
        \includegraphics[width=0.55\textwidth]{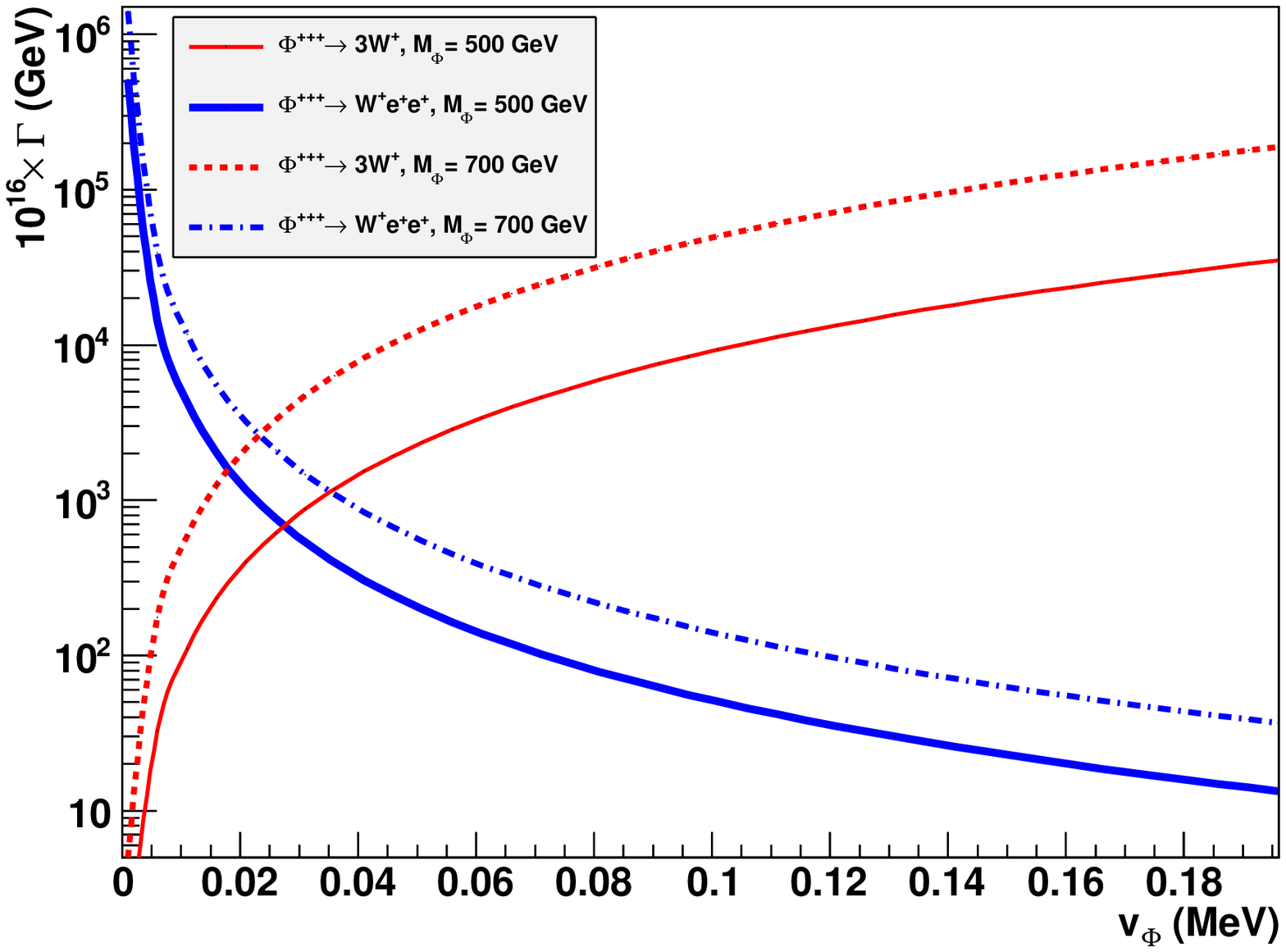}\caption{Decay widths as functions of $v_\Phi$.}
\end{tabular*}
\end{center}
\vspace{-0.75cm}
\end{figure}
The decay widths into the two modes are shown in Figs.~3 and 4. Note
the crossing point is $v_{\Phi} \sim 0.02 \textrm{--} 0.03\textrm{
MeV}$. For $M_{\Phi} = 500\textrm{ GeV}$, $\Gamma < 10^{-12}$ -- $6 \times 10^{-14}\textrm{ GeV}$, and we get displaced vertices.
For lower masses, widths are even smaller;  $\Phi^{+++}$ can escape
the detector!  For $v_{\Phi}> 0.2\textrm{ MeV}$, $\Phi^{+++}$ will
immediately decay to $W^{+} W^{+} W^{+}$.


\textbf{The SM Higgs mass:}  The $\Phi$ multiplet with tiny VEV
essentially behaves like an inert Higgs \cite{Barbieri:2006dq}. The SM Higgs mass can be
raised to $\sim 400 \textrm{--} 500\textrm{ GeV}$, if $v_\Phi$ is
large ($\sim\textrm{few --}38\textrm{ GeV}$). In that case, $H
\rightarrow \Phi^{+++} \Phi^{---}$.

\textbf{Neutrino mass hierarchy:}  If the mass of $\Phi^{+++}$ is
less than $3m_W$, then $\Phi^{+++} \rightarrow W^+ l^+l^+$
dominates. This will give rise to $ee$, $e\mu$, $\mu\mu$, along with
$\tau$'s. Dominance of $\mu\mu$ will indicate the normal hierarchy
for the light neutrino masses, while the dominance of $e\mu$ and
$ee$ will indicate the inverted hierarchy.

\vspace{-0.25cm}
\section{Conclusions}
\vspace{-0.25cm}
We have presented a new mechanism for neutrino mass generation with
a new scale at the TeV. The model links neutrino physics with the
physics of the TeV scale to be explored, and can be tested at the
LHC.
%

\vspace{-0.25cm}
\acknowledgments\vspace{-0.25cm}
The work presented in this talk was done in collaboration with K. S.
Babu and Z. Tavartkiladze. This work was supported in part by the U.
S. Department of Energy, Grant Numbers DE-FG02-04ER41306 and
DE-FG-02-04ER46140.


\vspace{-0.25cm}
\begin{thebibliography}{99}
\bibitem{seesaw}
P.~Minkowski,
  Phys.\ Lett.\ B {\bf 67} (1977) 421;
  M.~Gell-Mann, P.~Ramond and R.~Slansky, in {it Supergravity} eds.
  P. van Nieuwenhuizen and D.Z. Freedman (North Holland, Amsterdam, 1979) p. 315;
  T.~Yanagida,
{\it In Proceedings of the Workshop on the Baryon Number of the
Universe and Unified Theories, Tsukuba, Japan, 13-14 Feb 1979};
S.~L.~Glashow,
  NATO Adv.\ Study Inst.\ Ser.\ B Phys.\  {\bf 59} 687 (1979);
R.~N.~Mohapatra and G.~Senjanovic,
  Phys.\ Rev.\ Lett.\  {\bf 44} 912 (1980);  J.~Schechter and J.~W.~F.~Valle,
  Phys.\ Rev.\  D {\bf 22} 2227 (1980).

\bibitem{loop}
  A.~Zee,
  Phys.\ Lett.\  B {\bf 93}, 389 (1980);
  K.~S.~Babu,
  Phys.\ Lett.\  B {\bf 203}, 132 (1988);

 \bibitem{gn}
  S.~Gabriel and S.~Nandi,
  Phys.\ Lett.\  B {\bf 655}, 141 (2007)

 \bibitem{zurab}
 Z.~Tavartkiladze,
  Phys.\ Lett.\  B {\bf 528}, 97 (2002).

\bibitem{Amsler:2008zzb}
  C.~Amsler {\it et al.}  [Particle Data Group],
  Phys.\ Lett.\  B {\bf 667}, 1 (2008).

\bibitem{lep}
    LEP SUSY Working Group,LEPSUSYWG/02-05.1, http://lepsusy.web.cern.ch/lepsusy/Welcome.html0.

\bibitem{tevatron}
 CDF Collaboration, Aaltonen et al, arXiv 0902.1255 [hep-ex]; D0
Collaboration, V. M Abazov et al, Phys. ReV. Lett. 102, 161802
(2009).

\bibitem{Barbieri:2006dq}
  R.~Barbieri, L.~J.~Hall and V.~S.~Rychkov,
  Phys.\ Rev.\  D {\bf 74}, 015007 (2006).






\end{thebibliography}
\end{document}